\documentclass[aps,pra,showpacs,superscriptaddress,preprint]{revtex4}
\usepackage{amssymb}
\usepackage{amsmath}
\usepackage{graphicx}

\catcode`ð=\active
 \defð{\u{g}}
 \catcode`Ð=\active
 \defÐ{\u{G}}
 \catcode`Ý=\active
\defÝ{\. I}
 \catcode`ö=\active
\defö{\"{o}}
 \catcode`Ö=\active
 \defÖ{\"O}
 \catcode`ü=\active
 \defü{\"{u}}
 \catcode`Ü=\active
 \defÜ{\"{U}}
 \catcode`Þ=\active
\defÞ{\c{S}}
 \catcode`þ=\active
 \defþ{\c{s}}
 \catcode`ý=\active
 \defý{{\i}}
 \catcode`ç=\active
\defç{\d{c}}
 \catcode`Ç=\active
\defÇ{\d{C}}

\begin{document}

\title{\textbf{Exactly solvable effective mass }$D$-\textbf{dimensional Schr%
\"{o}dinger equation for pseudoharmonic and modified Kratzer problems }}
\author{Sameer M. Ikhdair}
\email[E-mail: ]{sikhdair@neu.edu.tr}
\affiliation{Department of Physics, Near East University, Nicosia, North Cyprus, Turkey}
\author{Ramazan Sever}
\email[E-mail: ]{sever@metu.edu.tr}
\affiliation{Department of Physics, Middle East Technical University, 06800, Ankara,Turkey}
\date{%
\today%
}

\begin{abstract}
We employ the point canonical transformation (PCT) to solve the $D$%
-dimensional Schr\"{o}dinger equation with position-dependent effective mass
(PDEM) function for two molecular pseudoharmonic and modified Kratzer
(Mie-type) potentials. In mapping the transformed exactly solvable $D$%
-dimensional ($D\geq 2$) Schr\"{o}dinger equation with constant mass into
the effective mass equation by employing a proper transformation, the exact
bound state solutions including the energy eigenvalues and corresponding
wave functions are derived. The well-known pseudoharmonic and modified
Kratzer exact eigenstates of various dimensionality is manifested.

Keywords: Bound states, point canonical transformation, position dependent
effective mass function, pseudoharmonic potential, modified Kratzer potential
\end{abstract}

\pacs{03.65.-w}
\maketitle

\newpage

\section{Introduction}

The solution of the Schr\"{o}dinger equation with position-dependent
effective mass (PDEM) for an arbitrary central potential has attracted
attention over the past years (cf., e.g. [1] and the references therein).
The motivation in this direction arises from considerable applications in
the different fields of the material science and condensed matter physics.
For instance, such applications in the case of the bound states in quantum
system [2], the nonrelativistic Green's function for quantum systems with
the position-dependent mass [3], the Dirac equation with position-dependent
mass in the Coulomb field [3], electronic properties of semiconductors [4], $%
^{3}He$ cluster [5], quantum dots [6], quantum liquids [7], graded alloys
and semiconductor heterostructures [8,9], the dependence of energy gap on
magnetic field in semiconductor nano-scale quantum rings [10], the solid
state problems with the Dirac equation [11], etc. Almost all of those works
mentioned above were focused on obtaining the energy eigenvalues and the
potential function for the given quantum system with the PDEM function. The
wave functions were either obtained by the solutions to the Schr\"{o}dinger
equation with the constant mass, or a few lower excited states were obtained
by acting of the creation operator on the ground state. The effective
potentials are the sum of the real potential form and the modification terms
emerged from the location dependence of the effective mass [12]. Taking into
consideration the PDEM, the main concern, in this work, is in obtaining the
energy spectra and/or wavefunctions of the $D$-dimensional Schr\"{o}dinger
equation with a given PDEM for central potentials by the point canonical
transformation (PCT) approach [1]. In modern theory, high dimensions are
also of interests in many fields [13-25].

In this work, we employ the PCT to solve the $D$-dimensional Schr\"{o}dinger
equation with PDEM for the pseudoharmonic [26-31] and modified Kratzer
[32-37] potentials through mapping this wave equation into the well-known
exactly solvable $D$-dimensional Schr\"{o}dinger equation with constant mass
for a given PDEM function [1]. Indeed, the PCT approach has enabled us to
obtain the exact effective mass bound state solutions including the energy
spectrum and corresponding wave functions in any dimension for the exactly
solvable classes of quantum molecular potentials.

This work is organized as follows: in section 2, we introduce the
methodology. Section 3 is mainly devoted to obtain the exact bound state
energy eigenvalue and eigen function solutions of the $D$-dimensional Schr%
\"{o}dinger equation with a given PDEM function for two exactly solvable
diatomic molecular potentials. We give a few concluding remarks in section 4.

\section{Methodology}

The $D$-dimensional PDEM Schr\"{o}dinger equation for a central potential $%
V(r)$ takes the form%
\begin{equation}
\overrightarrow{\mathbf{\nabla }}_{D}\left( \frac{1}{m(r)}\overrightarrow{%
\mathbf{\nabla }}_{D\text{ }}\psi _{l_{1}\cdots l_{D-2}}^{(l_{D-1}=l)}(%
\mathbf{x})\right) +2\left[ E-V(r)\right] \psi _{l_{1}\cdots
l_{D-2}}^{(l_{D-1}=l)}(\mathbf{x})=0,\text{ }
\end{equation}%
where the position-dependent mass distribution $m(r)$ is a real function.
Here the wave functions $\psi _{l_{1}\cdots l_{D-2}}^{(l_{D-1}=l)}(\mathbf{x}%
)$ belong to the energy eigenvalues $E$ and $V(r)$ stands for the $D$%
-dimensional standard central potential in the configuration space
coordinates. Here $r$ represents the $D$-dimensional radius $\left(
\sum_{i=1}^{D}x_{i}^{2}\right) ^{1/2}.$ Atomic units will be used
throughout, with $h/2\pi =\hbar =m_{0}=e=1$. Going over to a spherical
coordinate system with $D-1$ angular variables and one radial coordinate we
can write [21-25]%
\begin{equation}
\psi _{l_{1}\cdots l_{D-2}}^{(l)}(\mathbf{x})=r^{-(D-1)/2}R_{l}(r)Y_{l_{1}%
\cdots l_{D-2}}^{(l)}(\widehat{\mathbf{x}}),
\end{equation}%
where $Y_{l_{1}\cdots l_{D-2}}^{(l)}(\widehat{\mathbf{x}})$ represents
contribution from the hyperspherical harmonics that arise in higher
dimensions with $\mathbf{x}$ represents the $D$-dimensional position vector.
With the substitutions%
\begin{equation*}
\overrightarrow{\mathbf{\nabla }}_{D}\left( \frac{1}{m(r)}\overrightarrow{%
\mathbf{\nabla }}_{D\text{ }}\psi _{l_{1}\cdots l_{D-2}}^{(l_{D-1}=l)}(%
\mathbf{x})\right) =\left( \overrightarrow{\mathbf{\nabla }}_{D}\frac{1}{m(r)%
}\right) \cdot \left( \overrightarrow{\mathbf{\nabla }}_{D\text{ }}\psi
_{l_{1}\cdots l_{D-2}}^{(l_{D-1}=l)}(\mathbf{x})\right) +\frac{1}{m(r)}%
\mathbf{\nabla }_{D\text{ }}^{2}\psi _{l_{1}\cdots l_{D-2}}^{(l_{D-1}=l)}(%
\mathbf{x}),
\end{equation*}%
\begin{equation*}
\nabla _{D}^{2}=\frac{\partial ^{2}}{\partial r^{2}}+\frac{(D-1)}{r}\frac{%
\partial }{\partial r}-\frac{l(l+D-2)}{r^{2}},
\end{equation*}%
\begin{equation}
l(l+D-2)=-\left[ \frac{1}{\sin ^{D-2}\theta _{D-1}}\frac{\partial }{\partial
\theta _{D-1}}\left( \sin ^{D-2}\theta _{D-1}\frac{\partial }{\partial
\theta _{D-1}}\right) -\frac{L_{D-2}^{2}}{\sin ^{2}\theta _{D-1}}\right]
\end{equation}%
and the wave functions (2) into Eq. (1) will result in the following time
independent $D$-dimensional PDEM $l_{D}$th partial wave radial Schr\"{o}%
dinger wave equation%
\begin{equation}
\left\{ \frac{d^{2}}{dr^{2}}+\frac{m^{\prime }(r)}{m(r)}\left( \frac{D-1}{2}%
\frac{1}{r}-\frac{d}{dr}\right) -\frac{l_{D}\left( l_{D}+1\right) }{r^{2}}%
+2m(r)\left[ E-V(r)\right] \right\} R_{n,l}(r)=0,
\end{equation}%
where $m^{\prime }(r)=dm(r)/dr$ and the transformation $l\rightarrow
l_{D}=l+(D-3)/2,$ $D\geq 2$ and $\nabla _{D}^{2}$ stand for the $D$%
-dimensional angular momentum and Laplacian, respectively$.$ Moreover, the $%
D=1$ can be obtained through inserting $l=1$ and $0$ or $l_{D}=0$ and $-1.$
In the present work, we are concerned in bound states, i.e., $E<0.$ We also
should be careful about the behavior of the wave function $R(r)$ near $r=0$
and $r\rightarrow \infty .$ It may be mentioned that $R_{l}(r)$ should be
normalizable [8]. Obviously, it can be shown, most readily, that for a
constant mass, i.e., $m^{\prime }(r)=0$ case$,$ the above equation with
potential function $\widetilde{V}(s),$ angular momentum $\Lambda (l)$ and
energy spectrum $\widetilde{E}$ reduces to the usual case of
independent-position mass: [19,21-23]:
\begin{equation}
\left\{ \frac{d^{2}}{ds^{2}}-\frac{\Lambda _{D}(l)\left( \Lambda
_{D}(l)+1\right) }{s^{2}}+2\left[ \widetilde{E}-\widetilde{V}(s)\right]
\right\} \widetilde{R}_{n,\Lambda (l)}(s)=0,
\end{equation}%
where $\Lambda _{D}(l)=\Lambda (l)+(D-3)/2$. Consequently, the solutions for
a particular central potential $\widetilde{V}(s)$ are the same as long as $%
D+2\Lambda (l)$ remains unaltered. For example, the $s$-wave eigensolution $(%
\widetilde{R}_{0})$ and energy spectrum $(\widetilde{E})$ in
four-dimensional solutions are identical to the $p$-wave two-dimensional
solutions ($\Lambda (l)=0,D=4$)$\longrightarrow $($\Lambda (l)=1,D=2$). For
more detail on inter-dimensional degeneracies the reader may refer to, e.g.
[19,23,38,39]. We apply the following point canonical transformation (PCT) $%
s\rightarrow r$ (i.e., mapping function $s=q(r))$ with the substitution of
the wave function
\begin{equation}
R_{n,l}(r)=\frac{\widetilde{R}_{n,\Lambda (l)}(s)}{g(r)},
\end{equation}%
into Eq. (5) and after some simple algebra, the transformed Schr\"{o}dinger
equation becomes
\begin{equation*}
\left\{ \frac{d^{2}}{dr^{2}}+\left( 2\frac{g^{\prime }}{g}-\frac{q^{\prime
}{}^{\prime }}{q^{\prime }}\right) \frac{d}{dr}+\left( \frac{g^{\prime
}{}^{\prime }}{g}-\frac{q^{\prime }{}^{\prime }}{q^{\prime }}\frac{g^{\prime
}}{g}\right) \right.
\end{equation*}%
\begin{equation}
-\left. \Lambda _{D}(l)\left( \Lambda _{D}(l)+1\right) \left( \frac{%
q^{\prime }}{q}\right) ^{2}+2\left( q^{\prime }\right) ^{2}\left[ \widetilde{%
E}-\widetilde{V}(q(r))\right] \right\} R_{n,\Lambda (l)}(r)=0,
\end{equation}%
where the primes denote differentiation with respect to $r.$\ Further,
comparing Eq. (7) with Eq. (4), we find the following PCT transformations:

\begin{equation}
\frac{1}{2m(r)}\left( \frac{q^{\prime }{}^{\prime }}{q^{\prime }}-2\frac{%
g^{\prime }}{g}-\frac{m^{\prime }(r)}{m(r)}\right) =0,\text{ \ }q^{\prime
}=m(r)g^{2}(r)
\end{equation}%
\begin{equation}
V(r)=\frac{\left( q^{\prime }\right) ^{2}}{m(r)}\widetilde{V}(s),
\end{equation}%
\begin{equation}
E_{n}=\frac{\left( q^{\prime }\right) ^{2}}{m(r)}\widetilde{E}_{n},
\end{equation}%
and%
\begin{equation}
\Lambda _{D}(l)\left( \Lambda _{D}(l)+1\right) \left( \frac{q^{\prime }}{q}%
\right) ^{2}=\frac{l_{D}\left( l_{D}+1\right) }{r^{2}}-\frac{D-1}{2r}\frac{%
m^{\prime }(r)}{m(r)}-\frac{1}{2}\left[ F(m(r))-F(q^{\prime })\right] ,
\end{equation}%
where $F(x)=\frac{x^{\prime }{}^{\prime }}{x}-\frac{3}{2}\left( \frac{%
x^{\prime }}{x}\right) .$ Therefore, Eq. (6) and Eqs. (8)-(11) can be also
found for any potential system with PDEM.

\section{Applications}

We solve the $D$-dimensional PDEM Schr\"{o}dinger equation\ exactly for two
potentials: the pseudoharmonic potential [26-31] and the modified Kratzer
molecular potential [32-37]. The transformation function $g(r)$ will be
found for the given effective mass function $m(r)=m_{0}r^{\lambda }$ and the
selected PCT function $q(r)=r^{\nu }.$

\subsection{ pseudoharmonic potential}

The pseudoharmonic potential is given by [26-31]
\begin{equation}
\widetilde{V}(s)=V_{e}\left( \frac{s}{r_{e}}-\frac{r_{e}}{s}\right) ^{2},
\end{equation}%
as a reference potential where $V_{e}=\frac{1}{8}\kappa r_{e}^{2}$ is the
dissociation energy between two atoms in a solid with $\kappa $ is the force
constant and $r_{e}$ is the equilibrium bond length. The eigenvalue problem
(5) for\ $\widetilde{V}(s)$ in (12) can be solved analytically to get the
exact $D$-dimensional results for energy eigenvalues and wavefunctions of
this system (in units in which $m_{0}=\hbar =1)$ as [26-30]:%
\begin{equation}
\widetilde{E}_{n,\Lambda (l)}=-2V_{e}+\sqrt{\frac{V_{e}}{2r_{e}^{2}}}\left(
4n+2+\sqrt{(D+2\Lambda (l)-2)^{2}+8V_{e}r_{e}^{2}}\right) ,
\end{equation}%
and%
\begin{equation*}
\widetilde{R}_{n,\Lambda (l)}(s)=A_{l}s^{\sqrt{(\Lambda
(l)+D/2-1)^{2}+2V_{e}r_{e}^{2}}+1/2}\exp \left( -\frac{1}{2}\sqrt{\frac{%
2V_{e}}{r_{e}^{2}}}s^{2}\right)
\end{equation*}%
\begin{equation}
\times F\left( -n,\sqrt{(\Lambda (l)+D/2-1)^{2}+2V_{e}r_{e}^{2}}+1;\sqrt{%
\frac{2V_{e}}{r_{e}^{2}}}s^{2}\right) ,
\end{equation}%
with%
\begin{equation*}
A_{n,l}=\sqrt{\frac{2\left( \sqrt{2V_{e}}/r_{e}\right) ^{\sqrt{(\Lambda
(l)+D/2-1)^{2}+2V_{e}r_{e}^{2}}+1}n!}{\Gamma (n+\sqrt{(\Lambda
(l)+D/2-1)^{2}+2V_{e}r_{e}^{2}}+1)}}
\end{equation*}%
where $n=0,1,2,\cdots $ and $l=0,1,2,\cdots $ signify the usual radial and
angular momentum quantum numbers, respectively and $A_{n,l}$ being the
normalization constant. Here $F(-n,k+1,x)$ is a confluent hypergeometric
function. We follow Ref. [1] by taking the power law PDEM function $%
m(r)=m_{0}r^{\lambda }$ and the PCT function $q(r)=r^{\nu },$ where $m_{0}$
is the rest mass and $\lambda $ and $\nu $ are two non-zero real parameters.
We consider only the case where $\left( q^{\prime }\right) ^{2}/m(r)$ is
constant for which $\nu =1+\lambda /2,$ $\lambda \neq 2$ to avoid
position-dependent energy$.$ Inserting them into Eq. (6) and Eqs. (9)-(11),
we obtain%
\begin{equation}
V(r)=\frac{m_{0}}{2}\left( r^{1+\lambda /2}-\frac{r_{e}^{2}}{r^{1+\lambda /2}%
}\right) ^{2}C^{2},
\end{equation}%
\begin{equation}
E_{n,l}=\frac{2+\lambda }{2}\left[ -\eta r_{e}^{2}+1+2n+\sqrt{(\Lambda
(l)+D/2-1)^{2}+\eta ^{2}r_{e}^{4}}\right] C,
\end{equation}%
and%
\begin{equation*}
R_{n,l}(r)=a_{l}\left( \eta r\right) ^{\left( 1+\frac{\lambda }{2}\right)
\left( \sqrt{(\Lambda (l)+D/2-1)^{2}+\eta ^{2}r_{e}^{4}}+\frac{1}{2}\right) +%
\frac{\lambda }{4}}\exp \left( -\frac{\eta }{2}r^{2+\lambda }\right)
\end{equation*}%
\begin{equation}
\times F\left( -n,\sqrt{(\Lambda (l)+D/2-1)^{2}+\eta ^{2}r_{e}^{4}}+1;\eta
r^{2+\lambda }\right) ,
\end{equation}%
where $\eta =\sqrt{\kappa }/2=2m_{0}C/(2+\lambda )$ with $C=\nu \eta /m_{0}$
is a real potential parameter and $\Lambda (l)$ is%
\begin{equation}
\Lambda (l)=-\frac{\left( D-2\right) }{2}+\frac{1}{\left( 2+\lambda \right) }%
\sqrt{\left( D+2l-2\right) ^{2}+\left( 2+\lambda \right) ^{2}-2\left(
2+\lambda D\right) }.
\end{equation}%
The radial wave function (17) must vanish as $r\rightarrow 0$ and $%
r\rightarrow \infty .$ For the trivial case where $\lambda =0,$ i.e.,
constant mass ($m=m_{0}),$ we obtain $\Lambda (l)=l$ from Eq. (18). Hence,
we find out that Eqs. (15)-(17) agree with Eqs. (12)-(14). The same
procedure leads to the other exact solvable classes belong to various values
of parameter $\lambda .$ For example, if we insert $\lambda =2,$ the
following potential function with its energy spectrum and wave functions are
obtained%
\begin{equation}
V(r)=\frac{2\eta ^{2}}{m_{0}}\left( r^{2}-\frac{r_{e}^{2}}{r^{2}}\right)
^{2},
\end{equation}%
\begin{equation}
E_{n,l}=\frac{4\eta }{m_{0}}\left[ -\eta r_{e}^{2}+1+2n+\sqrt{(\Lambda
_{1}(l)+D/2-1)^{2}+\eta ^{2}r_{e}^{4}}\right] ,
\end{equation}%
and%
\begin{equation*}
R_{n,l}(r)=b_{l}\left( \eta r\right) ^{\sqrt{(2\Lambda
_{1}(l)+D-2)^{2}+4\eta ^{2}r_{e}^{4}}+\frac{3}{2}}\exp \left( -\frac{\eta }{2%
}r^{4}\right)
\end{equation*}%
\begin{equation}
\times F\left( -n,\sqrt{(\Lambda (l)+D/2-1)^{2}+\eta ^{2}r_{e}^{4}}+1;\eta
r^{4}\right) ,
\end{equation}%
with%
\begin{equation}
\Lambda _{1}(l)=-\frac{\left( D-2\right) }{2}+\frac{1}{4}\sqrt{\left(
D+2l-2\right) ^{2}-4\left( D-3\right) },
\end{equation}%
where $n,l=0,1,2,\cdots .$ The pseudoharmonic potential can be treated
exactly in three as well as in one and two dimensions. We note the special
cases $D=1,2$ and $3.$ For $D=2,$ with the customary notation $l=M$ and $%
r=\rho :$%
\begin{equation}
E_{n,M}=\frac{\eta \nu ^{2}}{m_{0}}\left[ -\eta r_{e}^{2}+1+2n+\sqrt{\frac{%
4M^{2}+\lambda ^{2}}{\left( 2+\lambda \right) ^{2}}+\eta ^{2}r_{e}^{4}}%
\right] ,
\end{equation}%
and%
\begin{equation*}
R_{n,M}(\rho )=a_{l}\left( \eta \rho \right) ^{\left( 1+\frac{\lambda }{2}%
\right) \left( \sqrt{\frac{4M^{2}+\lambda ^{2}}{\left( 2+\lambda \right) ^{2}%
}+\eta ^{2}r_{e}^{4}}+\frac{1}{2}\right) +\frac{\lambda }{4}}\exp \left( -%
\frac{\eta }{2}\rho ^{2+\lambda }\right)
\end{equation*}%
\begin{equation}
F\left( -n,\sqrt{\frac{4M^{2}+\lambda ^{2}}{\left( 2+\lambda \right) ^{2}}%
+\eta ^{2}r_{e}^{4}}+1;\eta \rho ^{2+\lambda }\right) \text{ }%
(n,M=0,1,2,\cdots ).
\end{equation}%
Furthermore, for constant mass case ($\lambda =0$):%
\begin{equation}
E_{n,M}=\frac{\eta }{m_{0}}\left[ -\eta r_{e}^{2}+1+2n+\sqrt{M^{2}+\eta
^{2}r_{e}^{4}}\right] ,
\end{equation}%
and%
\begin{equation}
R_{n,M}(r)=a_{l}\left( \eta r\right) ^{\left( \sqrt{M^{2}+\eta ^{2}r_{e}^{4}}%
+\frac{1}{2}\right) }\exp \left( -\frac{\eta }{2}r^{2}\right) F\left( -n,%
\sqrt{M^{2}+\eta ^{2}r_{e}^{4}}+1;\eta r^{2}\right) ,
\end{equation}%
which are identical to with those given in Ref. [28]. For $D=3:$
\begin{equation}
E_{n,l}=\frac{\eta \nu ^{2}}{m_{0}}\left[ -\eta r_{e}^{2}+1+2n+\sqrt{\frac{%
\left( 2l+1\right) ^{2}+\lambda \left( \lambda -2\right) }{\left( 2+\lambda
\right) ^{2}}+\eta ^{2}r_{e}^{4}}\right] ,
\end{equation}%
and%
\begin{equation*}
R_{n,l}(r)=a_{l}\left( \eta r\right) ^{\left( 1+\frac{\lambda }{2}\right)
\left( \sqrt{\frac{\left( 2l+1\right) ^{2}+\lambda \left( \lambda -2\right)
}{\left( 2+\lambda \right) ^{2}}+\eta ^{2}r_{e}^{4}}+\frac{1}{2}\right) +%
\frac{\lambda }{4}}\exp \left( -\frac{\eta }{2}r^{2+\lambda }\right)
\end{equation*}%
\begin{equation}
\times F\left( -n,\sqrt{\frac{\left( 2l+1\right) ^{2}+\lambda \left( \lambda
-2\right) }{\left( 2+\lambda \right) ^{2}}+\eta ^{2}r_{e}^{4}}+1;\eta
r^{2+\lambda }\right) ,
\end{equation}%
where $n,l=0,1,2,\cdots $ and inserting $\lambda =0$:%
\begin{equation}
E_{n,l}=\frac{\eta }{m_{0}}\left[ -\eta r_{e}^{2}+1+2n+\sqrt{\left(
l+1/2\right) ^{2}+\eta ^{2}r_{e}^{4}}\right] ,
\end{equation}%
and%
\begin{equation}
R_{n,l}(r)=a_{l}\left( \eta r\right) ^{\sqrt{\left( l+1/2\right) ^{2}+\eta
^{2}r_{e}^{4}}+\frac{1}{2}}\exp \left( -\frac{\eta }{2}r^{2}\right) F\left(
-n,\sqrt{\left( l+1/2\right) ^{2}+\eta ^{2}r_{e}^{4}}+1;\eta r^{2}\right) ,
\end{equation}%
which are identical with those given in Refs. [26,27,33]$.$ For $D=1$ ($s$%
-wave$)$:%
\begin{equation}
E_{n}=\frac{\eta \nu ^{2}}{m_{0}}\left[ -\eta r_{e}^{2}+1+2n+\sqrt{\left(
\frac{1+\lambda }{2+\lambda }\right) ^{2}+\eta ^{2}r_{e}^{4}}\right] ,
\end{equation}%
and%
\begin{equation*}
R_{n}(x)=a_{l}\left( \eta x\right) ^{\left( 1+\frac{\lambda }{2}\right)
\left( \sqrt{\left( \frac{1+\lambda }{2+\lambda }\right) ^{2}+\eta
^{2}r_{e}^{4}}+\frac{1}{2}\right) +\frac{\lambda }{4}}\exp \left( -\frac{%
\eta }{2}x^{2+\lambda }\right)
\end{equation*}%
\begin{equation}
F\left( -n,\sqrt{\left( \frac{1+\lambda }{2+\lambda }\right) ^{2}+\eta
^{2}r_{e}^{4}}+1;\eta x^{2+\lambda }\right) ,
\end{equation}%
and inserting $\lambda =0:$%
\begin{equation}
E_{n}=\frac{\eta }{m_{0}}\left[ -\eta r_{e}^{2}+1+2n+\frac{1}{2}\sqrt{%
1+4\eta ^{2}r_{e}^{4}}\right] ,
\end{equation}%
and%
\begin{equation}
R_{n}(x)=a_{l}\left( \eta x\right) ^{\frac{1}{2}\left( \sqrt{1+\eta
^{2}r_{e}^{4}}+1\right) }\exp \left( -\frac{\eta }{2}x^{2}\right) F\left( -n,%
\frac{1}{2}\sqrt{1+4\eta ^{2}r_{e}^{4}}+1;\eta x^{2}\right) ,
\end{equation}%
where $n=0,1,2,\cdots .$

\subsection{modified Kratzer molecular potential}

The modified Kratzer potential is [32-37]%
\begin{equation}
\widetilde{V}(s)=V_{e}\left( \frac{s-r_{e}}{s}\right) ^{2},
\end{equation}%
as a reference potential. The eigenvalue problem (5) for\ $\widetilde{V}(s)$
in (35) can be solved analytically to get the exact $D$-dimensional results
for energy eigenvalues and wavefunctions of this system (in units in which $%
m_{0}=\hbar =1)$ as [32-37]:%
\begin{equation}
\widetilde{E}_{n,\Lambda (l)}=V_{e}-\frac{1}{2a^{2}\left( 1+2n+\sqrt{\left[
D+2\Lambda (l)-2\right] ^{2}+8V_{e}r_{e}^{2}}\right) ^{2}},\text{ }%
n=l=0,1,2,\cdots
\end{equation}%
and%
\begin{equation*}
\widetilde{R}_{n,\Lambda (l)}(s)=B_{l}s^{\frac{1}{2}\left( 1+\sqrt{\left[
D+2\Lambda (l)-2\right] ^{2}+8V_{e}r_{e}^{2}}\right) }\exp \left[ -\beta s%
\right]
\end{equation*}%
\begin{equation}
\times F\left( -n,1+\sqrt{\left[ D+2\Lambda (l)-2\right] ^{2}+8V_{e}r_{e}^{2}%
};2ks\right) ,
\end{equation}%
with $a=1/\left( 4V_{e}r_{e}\right) $ and $B_{l}$ is the normalization
factor. The wave number $k$ for the modified Kratzer (pseudo-Coulomb)
spectrum under consideration is%
\begin{equation}
k=\frac{1}{a\left( 1+2n+\sqrt{\left[ D+2\Lambda (l)-2\right]
^{2}+8V_{e}r_{e}^{2}}\right) }.
\end{equation}%
The PDEM function $m(r)=m_{0}r^{\lambda }$ and the PCT function $q(r)=r^{\nu
}$ are same as before. Inserting them into Eq. (6) and Eqs. (9)-(11), we
obtain%
\begin{equation}
V(r)=P\left( \frac{r^{1+\lambda /2}-r_{e}}{r^{1+\lambda /2}}\right) ^{2},
\end{equation}%
\begin{equation}
E_{n,l}=P-\frac{32m_{0}r_{e}^{2}P^{2}}{\left[ \left( 1+2n\right) \left(
2+\lambda \right) +\sqrt{\left( 2+\lambda \right) ^{2}\left[ D+2\Lambda (l)-2%
\right] ^{2}+32m_{0}r_{e}^{2}P}\right] ^{2}},
\end{equation}%
and%
\begin{equation*}
R_{n,l}(r)=b_{l}r^{\frac{1}{4}\left( 2+\lambda +\sqrt{\left( 2+\lambda
\right) ^{2}\left[ D+2\Lambda (l)-2\right] ^{2}+32m_{e}r_{e}^{2}P}\right) +%
\frac{\lambda }{4}}\exp \left( -\gamma r^{1+\frac{\lambda }{2}}\right)
\end{equation*}%
\begin{equation}
\times F\left( -n,1+\frac{1}{\left( 2+\lambda \right) }\sqrt{\left(
2+\lambda \right) ^{2}\left[ D+2\Lambda (l)-2\right] ^{2}+32m_{e}r_{e}^{2}P}%
;2\gamma r^{1+\frac{\lambda }{2}}\right) ,
\end{equation}%
where $P=\frac{\left( 2+\lambda \right) ^{2}}{4m_{0}}V_{e}$ is a real
potential parameter and%
\begin{equation*}
\gamma =\frac{16m_{0}r_{e}P}{\left( 2+\lambda \right) \left[ \left(
1+2n\right) \left( 2+\lambda \right) +\sqrt{\left( 2+\lambda \right) ^{2}%
\left[ D+2\Lambda (l)-2\right] ^{2}+32m_{0}r_{e}^{2}P}\right] },
\end{equation*}%
which are identical with those given in Refs. [32,34-36] when $\lambda $ is
set to zero. The new angular momentum $\Lambda (l)$ is as given in Eq. (18).
Particularly, setting $\lambda =0$ with $P=V_{e}/m_{0}$ into Eqs. (39)-(41)$%
, $we recover the constant mass Eqs. (35)-(37). This manifests the
generality of our solution for the mass function given by $%
m(r)=m_{0}r^{\lambda }.$ We note the special cases $D=1,2$ and $3.$ For $D=2,
$ with customary notation $l=M,$ we obtain%
\begin{equation}
E_{n,M}=P-\frac{32m_{0}r_{e}^{2}P^{2}}{\left[ \left( 1+2n\right) \left(
2+\lambda \right) +\sqrt{16M^{2}+32m_{0}r_{e}^{2}P+4\lambda ^{2}}\right] ^{2}%
},
\end{equation}%
and%
\begin{equation*}
R_{n,M}(r)=b_{l}r^{\frac{1}{2}\left( 1+\frac{\lambda }{2}\right) \left( 1+%
\sqrt{16M^{2}+8V_{e}r_{e}^{2}+4\lambda ^{2}}\right) +\frac{\lambda }{4}}\exp
\left( -\gamma _{1}r^{1+\frac{\lambda }{2}}\right)
\end{equation*}%
\begin{equation}
F\left( -n,1+\sqrt{16M^{2}+8V_{e}r_{e}^{2}+4\lambda ^{2}};2\gamma _{1}r^{1+%
\frac{\lambda }{2}}\right) ,
\end{equation}%
with%
\begin{equation}
\gamma _{1}=\frac{16m_{0}r_{e}P}{\left( 2+\lambda \right) ^{2}}\frac{1}{1+2n+%
\sqrt{16M^{2}+8V_{e}r_{e}^{2}+4\lambda ^{2}}},
\end{equation}%
wher $n,M=0,1,2,\cdots .$ Further, when $\lambda =0$%
\begin{equation}
E_{n,M}=P-\frac{8m_{0}r_{e}^{2}P^{2}}{\left[ 1+2n+\sqrt{%
4M^{2}+8m_{0}r_{e}^{2}P}\right] ^{2}},
\end{equation}%
\begin{equation}
R_{n,M}(r)=b_{l}r^{\frac{1}{2}+\sqrt{4M^{2}+2V_{e}r_{e}^{2}}}\exp \left(
-\gamma _{1}r\right) F\left( -n,1+2\sqrt{4M^{2}+2V_{e}r_{e}^{2}};2\gamma
_{1}r\right) ,
\end{equation}%
where $P=V_{e}/m_{0}$ is a real potential parameter and%
\begin{equation}
\gamma _{1}=\frac{4m_{0}r_{e}P}{1+2n+\sqrt{16M^{2}+8V_{e}r_{e}^{2}}},
\end{equation}%
which are consistent with those given in Ref. [28]. For $D=3:$%
\begin{equation}
E_{n,l}=P-\frac{32m_{0}r_{e}^{2}P^{2}}{\left[ \left( 1+2n\right) \left(
2+\lambda \right) +2\sqrt{\left( 1+2l\right) ^{2}+8m_{0}r_{e}^{2}P+\lambda
(\lambda -2)}\right] ^{2}},
\end{equation}%
and%
\begin{equation*}
R_{n,l}(r)=b_{l}r^{\left( 1+\frac{\lambda }{2}\right) \sqrt{\frac{\left(
1+2l\right) ^{2}+\lambda (\lambda -2)}{\left( 2+\lambda \right) ^{2}}%
+2V_{e}r_{e}^{2}}+\frac{\lambda }{4}}\exp \left( -\gamma _{2}r^{1+\frac{%
\lambda }{2}}\right)
\end{equation*}%
\begin{equation}
\times F\left( -n,2\sqrt{\frac{\left( 1+2l\right) ^{2}+\lambda (\lambda -2)}{%
\left( 2+\lambda \right) ^{2}}+2V_{e}r_{e}^{2}};2\gamma _{2}r^{1+\frac{%
\lambda }{2}}\right)
\end{equation}%
\begin{equation*}
\gamma _{2}=\frac{16m_{0}r_{e}P}{\left( 2+\lambda \right) ^{2}}\frac{1}{%
1+2n+2\sqrt{\frac{\left( 2l+1\right) ^{2}+\lambda (\lambda -2)}{\left(
2+\lambda \right) ^{2}}+2V_{e}r_{e}^{2}}},
\end{equation*}%
where $n,l=0,1,2,\cdots .$ Interestingly for $\lambda =0$ the above results
in (48)-(49) reproduce the well-known three-dimensional relations obtained
recently by Refs. [32,34-36,39]. The energy eigenvalues and wave functions
are given in the form [32,34-36,40]
\begin{equation}
E_{n,l}=P-\frac{8m_{0}r_{e}^{2}P^{2}}{\left( 1+2n+\sqrt{\left( 1+2l\right)
^{2}+8m_{0}r_{e}^{2}P}\right) ^{2}}
\end{equation}%
$,$%
\begin{equation}
R_{n,l}(r)=b_{l}r^{\frac{1}{2}\sqrt{8V_{e}r_{e}^{2}+\left( 2l+1\right) ^{2}}%
}\exp \left( -\gamma _{2}r\right) F\left( -n,\sqrt{\left( 2l+1\right)
^{2}+8V_{e}r_{e}^{2}};2\gamma _{2}r\right) ,
\end{equation}%
\begin{equation}
\gamma _{2}=\frac{4m_{0}r_{e}P}{1+2n+\sqrt{\left( 2l+1\right)
^{2}+8V_{e}r_{e}^{2}}}.
\end{equation}%
For $D=1$ ($s$-wave):%
\begin{equation}
E_{n}=P-\frac{32m_{0}r_{e}^{2}P^{2}}{\left[ \left( 1+2n\right) \left(
2+\lambda \right) +2\sqrt{(1+\lambda )^{2}+8m_{0}r_{e}^{2}P}\right] ^{2}},
\end{equation}%
and%
\begin{equation}
R_{n}(r)=b_{l}r^{\left( 1+\frac{\lambda }{2}\right) \sqrt{2V_{e}r_{e}^{2}+%
\frac{(1+\lambda )^{2}}{\left( 2+\lambda \right) ^{2}}}+\frac{\lambda }{4}%
}\exp \left( -\gamma _{3}r^{1+\frac{\lambda }{2}}\right) F\left( -n,1+2\sqrt{%
\frac{(1+\lambda )^{2}}{\left( 2+\lambda \right) ^{2}}+2V_{e}r_{e}^{2}}%
;2\gamma _{3}r^{1+\frac{\lambda }{2}}\right) ,
\end{equation}%
\begin{equation*}
\gamma _{3}=\frac{16m_{0}r_{e}P}{\left( 2+\lambda \right) ^{2}}\frac{1}{%
\left( 1+2n+2\sqrt{\frac{(1+\lambda )^{2}}{\left( 2+\lambda \right) ^{2}}%
+2V_{e}r_{e}^{2}}\right) },
\end{equation*}%
where $n=0,1,2,\cdots .$ When $\lambda =0:$%
\begin{equation}
E_{n}=P-\frac{8m_{0}r_{e}^{2}P^{2}}{\left( 1+2n+\sqrt{1+8m_{0}r_{e}^{2}P}%
\right) ^{2}},
\end{equation}%
and%
\begin{equation}
R_{n}(r)=b_{l}r^{\frac{1}{2}\sqrt{8V_{e}r_{e}^{2}+1}}\exp \left( -\gamma
_{3}r\right) F\left( -n,1+\sqrt{1+8V_{e}r_{e}^{2}};2\gamma _{3}r\right) ,
\end{equation}%
\begin{equation*}
\gamma _{3}=\frac{4m_{0}r_{e}P}{1+2n+\sqrt{1+8V_{e}r_{e}^{2}}}.
\end{equation*}

\section{Cocluding Remarks}

In this work, we have studied the exact PDEM Schr\"{o}dinger equation in $D$%
-dimension for two diatomic molecular potentials, namely, pseudoharmonic
potential and modified Kratzer potential. The exactly solvable constant mass
$D$-dimensional Schr\"{o}dinger equation has been transformed into the form
similar to the effective mass by means of a proper PCT function. The mapping
of the resulting transformed equation with the original effective mass
equation provide us the required energy spectrum and wave functions for the
potential system under consideration. We have applied this methodology to
obtain the energy eigenvalues and the corresponding eigenfunctions of the
modified Kratzer and the pseudoharmonic potentials. The exact bound state
solutions of the constant mass Schr\"{o}dinger equation for the
pseudoharmonic and modified Kratzer problems are recovered for various
dimensionality upon inserting $D=1,2$ and $3$ and $\lambda =0.$

\acknowledgments Work partially supported by the Scientific and
Technological Research Council of Turkey (T\"{U}B\.{I}TAK).

\newpage

{\normalsize 
}

\newpage

\end{document}